\documentclass[a4paper,conference]{IEEEtran}

\usepackage[utf8]{inputenc}
\usepackage[T1]{fontenc}
\usepackage{url}
\usepackage{ifthen}
\usepackage[cmex10]{amsmath} 
\usepackage{stfloats}
\usepackage{amsmath}
\usepackage{epsfig}
\usepackage{amssymb}
\usepackage{verbatim}
\usepackage{delarray}
\usepackage{graphics}
\usepackage{epic}

\newcommand{\A}{{\mathcal{A}}}
\newcommand{\B}{{\mathcal{B}}}
\newcommand{\CC}{{\mathcal{C}}}

\newcommand{\FF}{{\mathcal{F}}}

\newcommand{\RR}{{\mathcal{R}}}
\newcommand{\SSS}{{\mathcal{S}}}

\newcommand{\U}{{\mathcal{U}}}

\newcommand{\F}{{\mathbb{F}}}

\newcommand{\Z}{{\mathbb{Z}}}
\newcommand{\zerob}{{\mathbf 0}}

\newcommand{\ab}{{\mathbf a}}

\renewcommand{\sb}{{\mathbf s}}

\newcommand{\Bf}{{\mathfrak{B}}}

\newcommand{\ie}{{\em i.e., }}
\newcommand{\eg}{{\em e.g., }}

\newcommand{\openbox}{\leavevmode
     \hbox to.77778em{%
     \hfil\vrule
     \vbox to.675em{\hrule width.6em\vfil\hrule}%
     \vrule\hfil}}
\newcommand{\qed}{\hspace*{1cm}\hspace*{\fill}\openbox}

\begin{document}
\title{Unique Factorization and Controllability of Tail-Biting Trellis Realizations via \\ Controller Granule Decompositions}

\author{%
\IEEEauthorblockN{G. David Forney, Jr.}
\IEEEauthorblockA{Laboratory for Information and Decision Systems \\
Massachusetts Institute of Technology \\
Cambridge, MA 02139 USA\\
              Email: forney@mit.edu}
}

\date{}

\maketitle

\begin{abstract}  The Conti-Boston factorization theorem (CBFT) for linear tail-biting trellis realizations is extended to group realizations with a new and simpler proof, based on a controller granule decomposition of the behavior and known controllability results for group realizations.  Further controllability results are given; e.g., a trellis realization is controllable if and only if its top (controllability) granule is trivial.  
\end{abstract}

\section{Introduction}

Tail-biting trellis realizations are  the simplest class of realizations of codes on cyclic graphs.  Decoding is generally simpler than for conventional trellis realizations \cite{CFV99}.

Koetter and Vardy \cite{KV02, KV03} developed the foundations of the theory of linear tail-biting trellis realizations.  Their key result was a  factorization theorem (KVFT), which shows that every reduced realization has a factorization into elementary trellises.

Recently, Conti and Boston \cite{C14} have proved a stronger unique factorization theorem (CBFT): the behavior (``label code") of a reduced linear tail-biting trellis realization factors uniquely into quotient spaces of ``span subcodes."   This  work was the main stimulus for the work reported here.  

Our main result is a generalization of the CBFT to group  realizations, with a new proof that we feel is even simpler and more insightful.  \cite[Remark III.3]{C14} notes that such a generalization is not straightforward.  

In Section \ref{GD}, we introduce a granule decomposition along the lines of the controller granule decomposition of minimal conventional trellis realizations of Forney and Trott \cite{FT93, FT04}, and  the span subcode decomposition of \cite{C14}.

In Section \ref{SCUF}, using results of \cite{F13} on the controllability of group realizations, we show that this granule decomposition yields a unique factorization of a group trellis behavior $\Bf$.  We  develop other controllability properties not considered in \cite{C14};  \eg the trellis diagram of an uncontrollable group trellis realization is disconnected \cite{FGL12}.  We show  that  the controller canonical realization based on this factorization is one-to-one, minimal, and group-theoretic, but possibly  nonhomomorphic.

Our development uses only  elementary group theory, principally the \emph{fundamental theorem of homomorphisms} (FTH) and the \emph{correspondence theorem} (CT).  For a brief introduction to the necessary group theory and our notation, see \cite{F13}.

\subsection{Preliminaries}\label{Section 2}

A (tail-biting) trellis realization $\RR$ of length $n$ is defined by a  set  of  $n$ \emph{symbol alphabets} $\{\A_j, j \in \Z_n\}$, a  set  of  $n$ \emph{state alphabets}  $\{\SSS_j, j \in \Z_n\}$, and a  set   of  $n$ \emph{constraint codes} $\{\CC_j \subseteq \SSS_j \times \A_j \times \SSS_{j+1}, j \in \Z_n\}$, where  index arithmetic is in $\Z_n$;  \eg $\CC_{n-1} \subseteq \SSS_{n-1} \times \A_{n-1} \times \SSS_{0}$.  

The \emph{configuration universe} $\U = \prod_{j \in \Z_n} \CC_j$ is thus a subset of $\SSS \times \A \times \SSS$, where $\A = \prod_{j \in \Z_n} \A_j$  and $\SSS = \prod_{j \in \Z_n} \SSS_j$.

In a linear trellis realization, each symbol or state alphabet is a finite-dimensional vector space over some  field $\F$, and each $\CC_j$ is a subspace of  $\SSS_j \times \A_j \times \SSS_{j+1}$, so $\U$ is a subspace of $\SSS \times \A \times \SSS$.  (In \cite{KV03} and \cite{C14}, it is assumed that $\A_j = \F$ always.)  In a group trellis realization, each symbol or state alphabet is a finite abelian group, and each $\CC_j$ is a subgroup of  $\SSS_j \times \A_j \times \SSS_{j+1}$, so $\U$ is a subgroup of $\SSS \times \A \times \SSS$.

The \emph{extended behavior} $\bar{\Bf}$ of $\RR$ is the set of configurations $(\sb, \ab, \sb')  \in \U$ such that $\sb = \sb'$;  \ie such that  the constraints of $\U$ and the equality constraints $\sb = \sb'$ are both satisfied \cite{F13}.  Its \emph{behavior} $\Bf$ is the projection of $\bar{\Bf}$ onto $\A \times \SSS$, which is an isomorphism. The \emph{code} $\CC$ realized by $\RR$ is the projection of $\bar{\Bf}$ or $\Bf$  onto $\A$.

The (normal) \emph{graph} of $\RR$ \cite{F13} is the single-cycle graph with $n$ vertices corresponding to the constraint codes $\CC_j$, $n$ edges corresponding to the state variables $\SSS_j$, where edge $\SSS_j$ is incident on vertices $\CC_{j-1}$ and $\CC_j$, and $n$ half-edges corresponding to the symbol variables $\A_j$, where half-edge $\A_j$ is incident only on vertex $\CC_j$.

\section{Granule decomposition}\label{GD}

\subsection{Partial ordering of fragments} 

A \emph{proper fragment} of a trellis realization $\RR$ corresponds to a \emph{circular interval} $[j,k), j \in \Z_n, k \in \Z_n$, and will  be denoted by $\FF^{[j,k)}$.   $\FF^{[j,k)}$  includes the constraint codes $\{\CC_{j'}, j' \in [j,k)\}$ and the \emph{internal state variables} $\{\SSS_{j'}, j' \in (j,k)\}$, and has    \emph{boundary} $\{\SSS_j, \SSS_k\}$.   Accordingly, we define its \emph{vertex set} as $V(\FF^{[j,k)}) = [j,k)$, and its \emph{edge set} as $E(\FF^{[j,k)}) = (j,k)$.   The (normal) graph of every proper fragment is cycle-free.

We  define the \emph{level} of $\FF^{[j,k)}$ as the number $\ell = |E(\FF^{[j,k)})|$ of its internal state variables;  \ie $\ell = k-j-1$ mod $n$. 
Thus  $|V(\FF^{[j,k)})| = \ell+1.$   We may denote a level-$\ell$ fragment $\FF^{[j,j+\ell+1)}$ by  $\FF^{[j,j+\ell]}$. A level-$(n \! - \! 1)$ fragment $\FF^{[j,j)}$ is  obtained from $\RR$ by cutting the edge $\SSS_j$ into two half-edges;  it contains all $n$ constraint codes and $n-1$ internal state variables.   A level-0 fragment $\FF^{[j,j+1)} = \FF^{[j,j]}$ contains one constraint code $\CC_j$ and no internal state variables.

We also regard the entire realization $\RR$ as a fragment, whose level is $n$.  $\RR$ contains  $\ell = |E(\RR)| =  n$ internal state variables, and  $\ell =|V(\RR)| =  n$  (not  $\ell+1$) constraint codes.

As observed in \cite{C14},
the set $\mathfrak{F}(\RR)$ of  fragments of a tail-biting trellis realization $\RR$ is  partially ordered by set inclusion.  The maximum fragment $\RR$ includes all proper fragments $\FF^{[j,k)}$. The partial ordering of  proper fragments corresponds to the partial ordering of the circular intervals $[j,k)$ by set inclusion;  \ie $\FF^{[j',k')} \le \FF^{[j,k)}$ iff $[j',k') \subseteq [j,k)$. 
The minimal  fragments are the level-0 fragments $\FF^{[j,j+1)}$.

The partial ordering of $\mathfrak{F}(\RR)$ may be illustrated by a \emph{Hasse diagram}, as follows.  A fragment $\FF' \in \mathfrak{F}(\RR)$ is said to be \emph{covered} by another fragment $\FF \in\mathfrak{F}(\RR)$ if $\FF' < \FF$ and there is no fragment $\FF'' \in \mathfrak{F}(\RR)$ such that $\FF' < \FF'' < \FF$ \cite{Stanley}.  In our setting, $\FF'$ is  covered by $\FF$ if  $\FF' < \FF$ and the level of $\FF'$ is one less than the level of $\FF$.     The set $\mathfrak{F}(\RR)$ is  thus said to be \emph{graded} by  level (number of internal state variables).

\begin{figure}[h]
\setlength{\unitlength}{5pt}
\centering
\begin{picture}(28,20)(1, 1)
\put(0,0){$\FF^{[0,1)}$}
\put(10,0){$\FF^{[1,2)}$}
\put(20,0){$\FF^{[2,3)}$}
\put(30,0){$\FF^{[3,0)}$}
\put(0,5){$\FF^{[0,2)}$}
\put(10,5){$\FF^{[1,3)}$}
\put(20,5){$\FF^{[2,0)}$}
\put(30,5){$\FF^{[3,1)}$}
\put(0,10){$\FF^{[0,3)}$}
\put(10,10){$\FF^{[1,0)}$}
\put(20,10){$\FF^{[2,1)}$}
\put(30,10){$\FF^{[3,2)}$}
\put(0,15){$\FF^{[0,0)}$}
\put(10,15){$\FF^{[1,1)}$}
\put(20,15){$\FF^{[2,2)}$}
\put(30,15){$\FF^{[3,3)}$}
\put(3,2){\line(0,1){3}}
\put(12,2){\line(-3,1){9}}
\put(13,2){\line(0,1){3}}
\put(22,2){\line(-3,1){9}}
\put(23,2){\line(0,1){3}}
\put(32,2){\line(-3,1){9}}
\put(33,2){\line(0,1){3}}
\put(5.5,1.5){\line(6,1){24}}
\put(3,7){\line(0,1){3}}
\put(12,7){\line(-3,1){9}}
\put(13,7){\line(0,1){3}}
\put(22,7){\line(-3,1){9}}
\put(23,7){\line(0,1){3}}
\put(32,7){\line(-3,1){9}}
\put(33,7){\line(0,1){3}}
\put(5.5,6.5){\line(6,1){24}}
\put(3,12){\line(0,1){3}}
\put(12,12){\line(-3,1){9}}
\put(13,12){\line(0,1){3}}
\put(22,12){\line(-3,1){9}}
\put(23,12){\line(0,1){3}}
\put(32,12){\line(-3,1){9}}
\put(33,12){\line(0,1){3}}
\put(5.5,11.5){\line(6,1){24}}
\put(17,20.3){$\RR$}
\put(18,20){\line(1,-1){3}}
\put(17,20){\line(-1,-1){3}}
\put(19,20){\line(4,-1){11}}
\put(16,20){\line(-4,-1){11}}
\put(-8,0){$\ell = 0$}
\put(-8,5){$\ell = 1$}
\put(-8,10){$\ell = 2$}
\put(-8,15){$\ell = 3$}
\put(-8,20){$\ell = 4$}
\end{picture}
\caption{Hasse diagram of $\mathfrak{F}(\RR)$ when $n = 4$.}
\label{HDTBT}
\end{figure}
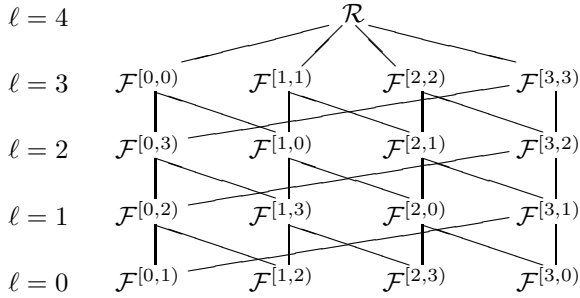

The Hasse diagram of $\mathfrak{F}(\RR)$ is illustrated in Figure \ref{HDTBT} for a tail-biting trellis realization $\RR$ of length $n=4$.  

As numerous authors have observed (\eg \cite{KV03, C14}), a conventional trellis realization may be viewed as a special case of a tail-biting trellis realization in which $\SSS_0$ is trivial.  Correspondingly, the Hasse diagram of a conventional trellis realization is a subdiagram of the Hasse diagram for a tail-biting trellis realization $\RR$ of the same length, comprising the fragments $\{\FF \in \mathfrak{F}(\RR) \mid \FF \le \FF^{[0,0)}\}$.   By cyclic rotation of the index set $\Z_n$, any level-$(n \! - \! 1)$ fragment $\FF^{[j,j)}$ may be regarded as a conventional trellis realization.

\subsection{Subbehaviors}\label{SFG}

For every proper fragment $\FF = \FF^{[j,k)} \in \mathfrak{F}(\RR)$, we define the \emph{subbehavior} $\Bf_{\FF} = \Bf^{[j,k)}$ as the set of $(\ab, \sb) \in \Bf$ that are all-zero on or outside the boundary of $\FF$.  For example, $\Bf^{[0,0)}$ is the behavior of a conventional trellis realization of length $n$. We also define $\Bf_\RR = \Bf$.

Evidently  if $\FF' \le \FF$, then $\Bf_{\FF'} \subseteq \Bf_{\FF}$.  Thus the set $\{\Bf_\FF, \FF \in \mathfrak{F}(\RR)\}$ has the same partial ordering as $\mathfrak{F}(\RR)$. 

For a level-0 fragment $\FF^{[j,j]}$, we have 
$$\Bf^{[j,j]} = \{(\ab, \zerob) \mid a_j \in (\CC_j)_{:\A_j}, a_{j'} = 0 \mathrm{~if~} j' \neq j\},$$
where  $(\CC_j)_{:\A_j} = \{a_j \in \A_j \mid (0, a_j, 0) \in \CC_j\}$ is the \emph{cross-section} of $\CC_j$ on $\A_j$.  As in \cite{F13},  $(\CC_j)_{:\A_j}$ will be denoted by $\underline{\A}_j$, and  called the \emph{nondynamical symbol alphabet} of $\CC_j$. 

\subsection{Granules}

For non-level-0 fragments,  we  define $\Bf_{< \FF}$ as the behavior generated by all $\Bf_{\FF'}$ such that $\FF' < \FF$, as in \cite{C14}. 
In other words, $\Bf_{< \FF} = \sum_{\FF' < \FF} \Bf_{\FF'}$.
Evidently $\Bf_{< \FF} \subseteq \Bf_\FF$.  

We  define the \emph{controller granule} $\Gamma_\FF$ as the quotient  $\Bf_\FF/\Bf_{< \FF}$.  In the linear case, $\Bf_\FF$ and $\Bf_{< \FF}$ are vector spaces, and their quotient $\Gamma_\FF$ is a vector space of dimension $\dim \Gamma_\FF = \dim \Bf_\FF - \dim \Bf_{<\FF}$.  In the group case, $|\Gamma_\FF| = |\Bf_\FF|/|\Bf_{<\FF}|$.

For a level-0 fragment $\FF^{[j,j+1)}$, we define the \emph{nondynamical granule} $\Gamma_\FF$ as $\Bf^{[j,j+1)} \cong \underline{\A}_j$.  The set $\{\Gamma_\FF, \FF \in \mathfrak{F}(\RR)\}$  thus consists of nondynamical granules at level $\ell  = 0$, and controller granules at levels $\ell > 0$.

At level $n$, where $\FF = \RR$,  we will call $\Gamma_\RR = \Bf/\Bf_{<\RR}$ the \emph{top granule} of $\RR$, or the \emph{controllability granule} of $\RR$, since as we will see $\Gamma_\RR$ governs the controllability properties of $\RR$.  

Note that $\Bf_{<\RR} = \sum_{j} \Bf^{[j,j)}$, the behavior generated by all level-$(n \! - \! 1)$ subbehaviors $\Bf^{[j,j)}$.  We will  call $\Bf_{<\RR}$ the \emph{controllable subbehavior} $\Bf^c$ of $\Bf$.

At levels $1 \le \ell \le n-1$, a proper fragment $\FF^{[j,k)}$   covers precisely two fragments, namely $\FF^{[j,k-1)}$ and $\FF^{[j+1,k)}$.  Thus
$\Bf_{<\FF^{[j,k)}} = \Bf^{[j,k-1)} + \Bf^{[j+1,k)}$, and the corresponding controller granule is 
$$\Gamma^{[j,k)} = \frac{\Bf^{[j,k)}}{\Bf^{[j,k-1)} + \Bf^{[j+1,k)}}.$$
Forney and Trott \cite{FT93, FT04} define a controller granule for a  conventional group trellis realization similarly as $\Gamma^{[j,k)} = \CC^{[j,k)}/(\CC^{[j,k-1)} + \CC^{[j+1,k)})$, where the subcode $\CC^{[j,k)} \subseteq \CC$ is the set of $\ab \in \CC$ that are all-zero outside the boundary of $\FF^{[j,k)}$.  The two definitions turn out to be equivalent for minimal conventional trellis realizations.

\subsection{$\ell$-controllable behaviors}

For $0 \le \ell \le n-1$,  we  define the \emph{$\ell$-controllable behavior} $\Bf_\ell$ as the behavior generated by all  level-$\ell$ subbehaviors $\Bf^{[j,j+\ell]}$. 
In other words, $\Bf_\ell = \sum_{j} \Bf^{[j,j+\ell]}.$  Note that $\Bf_{n-1} = \Bf^c$, the controllable subbehavior of $\Bf$.  We also define $\Bf_{n} = \Bf$.

Evidently $\Bf_{\ell-1} \subseteq \Bf_\ell$ for $1 \le \ell \le n$.
Moreover,  $\Bf_0 = \sum_{j}\Bf^{[j,j+1)} = \underline{\A} \times \{\zerob\}$, where $\underline{\A} =  \{\ab \in \A \mid (\ab, \zerob) \in \Bf\} = \prod_j \underline{\A}_j$. 
We call $\Bf_0$ the \emph{nondynamical behavior} of $\RR$.

We thus have a chain of subgroups $\Bf_0 = \underline{\A} \times \{\zerob\} \subseteq \Bf_1 \subseteq \cdots \subseteq \Bf_{n} = \Bf$, which is  a normal series since all groups are abelian.  We denote the factor groups of this chain by $Q_\ell = \Bf_\ell/\Bf_{\ell-1}, 1 \le \ell \le n$, plus $Q_0 = \Bf_0$.

By elementary group theory, we have $|\Bf| = \prod_{\ell}|Q_\ell|$;
or, in the linear case, $\dim \Bf  = \sum_\ell \dim Q_\ell$.  If we define sets $[Q_\ell]$ of  coset representatives  for the cosets of $\Bf_{\ell-1}$ in $\Bf_\ell$, then every $(\ab, \sb) \in \Bf$ may be uniquely expressed as a sum of coset representatives;  or, in the linear case, if we define a basis $\B_\ell$ for each quotient  $Q_\ell$, then every $(\ab, \sb) \in \Bf$ may be uniquely expressed as a  linear combination of basis elements.  

Since $Q_\ell$ is generated by the elements of $\Bf_\ell$ that are not in $\Bf_{\ell-1}$, and every  element of $\Bf_\ell$ is an element of some level-$\ell$ subbehavior $\Bf^{[j,j+\ell]}$, the nonzero coset representatives  in $[Q_\ell]$ may  all be taken as elements of  some $\Bf^{[j,j+\ell]} \setminus \Bf_{\ell - 1}$.
We note that if $(\ab, \sb) \in \Bf^{[j,j+\ell]} \setminus \Bf_{\ell - 1}$, then the support of $\sb$ must be precisely the length-$\ell$ circular interval $[j+1,j+\ell]$, else $(\ab, \sb) \in \Bf_{\ell-1}$.

The level-$\ell$ subbehaviors $\Bf^{[j,j+\ell]}$  thus comprise a sufficient set of representatives for $Q_\ell$.  We  say that  \emph{unique  factorization} holds if every element of every level-$\ell$ behavior $\Bf_\ell$ is a unique sum of elements of level-$\ell$ subbehaviors $\Bf^{[j,j+\ell]}$, modulo $\Bf_{\ell-1}$; \ie if $\Bf_\ell$ modulo $\Bf_{\ell-1}$ is the (internal) \emph{direct sum} 
$$\Bf_\ell = \bigoplus_{j \in \Z_n} \Bf^{[j,j+\ell]} \mod \Bf_{\ell-1}.$$

\section{Controllability and unique factorization}\label{SCUF}

In previous work \cite{FGL12, F13}, we have defined controllability as the property of  ``having independent constraints," since we have proved that a realization is observable if and only if its dual realization has this property.

We now  show that 
for a linear or group tail-biting trellis realization $\RR$, controllability in this sense is equivalent to the property that the top granule $\Gamma_\RR$ is trivial.  Simultaneously, we obtain an easy proof that unique factorization holds for $\RR$, under the proviso (as in  \cite{KV02, KV03, C14}) that $\RR$ is \emph{reduced};  that is, $\RR$ is \emph{state-trim}--- \ie $\Bf_{|\SSS_j} = \SSS_j$ for all $j$--- and $\RR$ is \emph{branch-trim}--- \ie $\Bf_{|\SSS_j \times \A_j \times \SSS_{j+1}} = \CC_j$ for all $j$.
\vspace{1ex}

(Notation:  in this section, we will use notation appropriate to the group case--- \ie we use sizes rather than dimensions;  the reader may translate to the linear case if desired.)

\subsection{Controllability}

In \cite{FGL12, F13}, a realization $\RR$ is called \emph{controllable} if the the constraints of $\U$ and the equality constraints $\sb = \sb'$ are independent.  More concretely, $\RR$ is controllable if the image $\SSS^c$  of the syndrome-former homomorphism $\U \to \SSS$ defined by $(\sb, \ab, \sb') \mapsto \sb - \sb'$ is equal to $\SSS$.  Since the kernel of this homomorphism is the extended behavior $\bar{\Bf}$, we  have $\U/\bar{\Bf} \cong \SSS^c \subseteq \SSS$, by the FTH.  This yields the following \textbf{controllability test}: 
$|\U|/|\bar{\Bf}|  \le |\SSS|,$
with equality if and only if $\RR$ is controllable \cite{F13}. 
In other words, since $\Bf \cong \bar{\Bf}$, a  realization is uncontrollable if and only if its  constraints are dependent in the following sense:\footnote{This result may be understood as follows.  Ignoring state equality constraints, there are  $|\U| = \prod_{j} |\CC_j|$ possible configurations.  If the state equality constraints $\{s_j = s'_j, j \in \Z_n\}$ are all independent of  the  set of code constraints $\{\CC_j, j \in \Z_n\}$, then each state equality constraint $s_j = s'_j$ reduces the number of possible configurations by a factor of $|\SSS_j|$, so $|\Bf| = |\U|/|\SSS|$, where $|\SSS| = \prod_{j} |\SSS_j|$.  If the constraints are dependent--- \ie if $\RR$ is not controllable--- then the reduction is strictly less, and $|\Bf| > |\U|/|\SSS|.$}
 $$|\Bf| > \frac{|\U|}{|\SSS|} = \frac{\prod_j |\CC_j|}{\prod_j |\SSS_j|}.$$

\subsection{Disconnected trellis realizations}\label{CTR}

We  now show that if the top granule $\Gamma_\RR = \Bf/\Bf^c$ is nontrivial, then $\Bf$ consists of $|\Gamma_\RR|$ disconnected subbehaviors, namely the cosets of the controllable subbehavior $\Bf^c = \sum_j \Bf^{[j,j)}$ in $\Bf$.  
Similar results were proved  in \cite{FGL12} and \cite[Appendix A]{GLF12};  the proof here is simpler, and does not rely on duality.

\vspace{1ex}
\noindent
\textbf{Lemma}.  For a linear or group trellis realization $\RR$ with behavior $\Bf$ and controllable subbehavior $\Bf^c$, for any $j \in \Z_n$: 

\vspace{1ex}
(a) $\Bf_{|\SSS_j}/(\Bf^c)_{|\SSS_j} \cong \Gamma_\RR;$  

 (b) $\Bf_{|\SSS_j \times \A_j \times \SSS_{j+1}}/(\Bf^c)_{|\SSS_j \times \A_j \times \SSS_{j+1}} \cong \Gamma_\RR.$

\vspace{1ex}
\noindent
\textit{Proof}.  (a) The projections of $\Bf$ and $\Bf^c$ onto $\SSS_j$ have a common kernel $\Bf^{[j,j)} = \{(\ab, \sb) \in \Bf \mid s_j = 0\}$.  Thus $\Bf_{|\SSS_j}/(\Bf^c)_{|\SSS_j} \cong \Bf/\Bf^c = \Gamma_\RR$, by the CT.

  (b) The projections of $\Bf$ and $\Bf^c$ onto $\SSS_j \times \A_j \times \SSS_{j+1}$ have a common kernel $\Bf^{[j+1,j)} = \{(\ab, \sb) \in \Bf \mid (s_j, a_j, s_{j+1}) = (0,0,0)\}$, so (b) follows also from the CT.
\qed \vspace{1ex}

If $\RR$ is reduced, as we assume, then $\Bf_{|\SSS_j} = \SSS_j$ and $\Bf_{|\SSS_j \times \A_j \times \SSS_{j+1}} = \CC_j$.  Moreover, we may regard  $\Bf^c$ as the behavior of the \emph{controllable subrealization} $\RR^c$ of $\RR$, defined as the reduced tail-biting trellis realization with state spaces $(\SSS_j)^c = (\Bf^c)_{|\SSS_j}$, symbol spaces $\A_j$, and constraint codes $(\CC_j)^c = (\Bf^c)_{|\SSS_j \times \A_j \times \SSS_{j+1}}$. This lemma then states that $\SSS_j/(\SSS_j)^c \cong \Gamma_\RR$ and $\CC_j/(\CC_j)^c \cong \Gamma_\RR$.
\vspace{0.5ex}

More concretely,  (a) implies that, if $\Gamma_\RR$ is nontrivial, then for each $j$, each coset $\Bf^c + (\ab, \sb)$ of $\Bf^c$ in $\Bf$ passes through a distinct corresponding  coset $(\SSS_j)^c + (\sb)_j$ of $(\SSS_j)^c$ in $\SSS_j$.  Similarly, $\CC_j$ partitions into $|\Gamma_\RR|$ disjoint cosets of $(\CC_j)^c$, each representing state transitions within one coset of $\Bf^c$ in $\Bf$.  The trellis diagram of $\RR$ thus consists of $|\Gamma_\RR|$ disconnected subdiagrams, one representing  each coset of $\Bf^c$ in $\Bf$.  Thus for any $j, j'$, there is no trajectory $(\ab, \sb)$ connecting any state $s_j$ in a given coset of $(\SSS_j)^c$ in $\SSS_j$ to a state  $s_{j'}$ in a coset of $(\SSS_{j'})^c$ in $\SSS_{j'}$, unless the two cosets correspond to the same coset of $\Bf^c$ in $\Bf$.

\subsection{First-state chain}\label{GI}

We now show that the controller granules of $\RR$ are isomorphic to factor groups of certain normal series.

\vspace{1ex}
\noindent
\textbf{Lemma} (\textbf{first-state chain}).  For $j \in \Z_n$, $1 \le \ell \le n-1$,  
$$ \Gamma^{[j,j+\ell]}\cong  \frac{(\Bf^{[j,j+\ell]})_{|\SSS_j \times \A_j \times \SSS_{j+1}}}{(\Bf^{[j,j+\ell)})_{|\SSS_j \times \A_j \times \SSS_{j+1}}}   \cong   \frac{(\Bf^{[j,j+\ell]})_{|\SSS_{j+1}}}{(\Bf^{[j,j+\ell)})_{|\SSS_{j+1}}} . $$

\vspace{1ex}
\noindent
\textit{Proof}. We have $\Gamma^{[j,j+\ell]} = \Bf^{[j,j+\ell]}/(\Bf^{[j,j+\ell)} + \Bf^{(j, j + \ell]})$.
The projections of $\Bf^{[j,j+\ell]}$ and $\Bf^{[j,j+\ell)} + \Bf^{(j, j + \ell]}$ onto $\SSS_j \times \A_j \times \SSS_{j+1}$ are $(\Bf^{[j,j+\ell]})_{|\SSS_j \times \A_j \times \SSS_{j+1}}$ and $(\Bf^{[j,j+\ell)})_{|\SSS_j \times \A_j \times \SSS_{j+1}}$, respectively, and their common kernel is $\Bf^{(j,j+\ell]} = \{(\ab, \sb) \in \Bf^{[j,j+\ell]} \mid (s_j, a_j, s_{j+1}) = (0,0,0)\}$. 
Similarly, the  projections of $(\Bf^{[j,j+\ell]})_{|\SSS_j \times \A_j \times \SSS_{j+1}}$ and $(\Bf^{[j,j+\ell)})_{|\SSS_j \times \A_j \times \SSS_{j+1}}$ onto $\SSS_{j+1}$ are  $(\Bf^{[j,j+\ell]})_{|\SSS_{j+1}}$ and  $(\Bf^{[j,j+\ell)})_{|\SSS_{j+1}}$, respectively, and their common kernel is $(\Bf^{[j,j]})_{|\SSS_j \times \A_j \times \SSS_{j+1}} = \{0\} \times \underline{\A}_j \times \{0\}.$
Thus both isomorphisms follow from the CT. \qed \vspace{1ex}

It follows from the first isomorphism that for each $\CC_j$ there  is a normal series
$
(\Bf^{[j,j]})_{|\SSS_j \times \A_j \times \SSS_{j+1}} = \{0\} \times \underline{\A}_j \times \{0\} \subseteq  (\Bf^{[j,j+1]})_{|\SSS_j \times \A_j \times \SSS_{j+1}} 
\subseteq \cdots \subseteq (\Bf^{[j,j)})_{|\SSS_j \times \A_j \times \SSS_{j+1}},$
whose factor groups are isomorphic to the granules $\Gamma^{[j,j+\ell]}$, $ 0 \le \ell \le n-1$.
This chain implies that $$|(\Bf^{[j,j)})_{|\SSS_j \times \A_j \times \SSS_{j+1}}| =   \prod_{\ell=0}^{n-1} |\Gamma^{[j,j+\ell]}|.$$  This result will be useful in the next section.

It follows from the second isomorphism that for each state space $\SSS_{j+1}$ there is a normal series
$
(\Bf^{[j,j]})_{|\SSS_{j+1}} = \{0\} \subseteq (\Bf^{[j,j+1]})_{|\SSS_{j+1}} \subseteq \cdots \subseteq (\Bf^{[j,j)})_{|\SSS_{j+1}},
$
whose factor groups are isomorphic to the granules $\Gamma^{[j,j+\ell]}, 1 \le \ell \le n-1$.  We call this normal series the \emph{first-state chain} at $\SSS_{j+1}$, since $\SSS_{j+1}$ is the first possibly nonzero state in the trajectories in $\Bf^{[j,j+\ell]}, 1 \le \ell \le n-1$.  This chain implies that $$|(\Bf^{[j,j)})_{|\SSS_{j+1}}| = \prod_{\ell = 1}^{n-1} |\Gamma^{[j,j+\ell]}|.$$

\vspace{-2ex}
\subsection{Controllability and unique factorization}\label{CUF}

We will now show that $\RR$ is controllable if and only if $\Bf = \Bf^c$;  \ie if and only if the top granule $\Gamma_\RR$ is trivial.  Moreover, the controller granule decomposition gives a unique factorization of both $\Bf^c$ and $\Bf$.

We first state a technical lemma that shows that in the controllable subrealization $\RR^c$, the number of transitions $(s_j, a_j, s_{j+1}) \in (\CC_j)^c$ is the number of states $s_j \in (\SSS_j)^c$ times the number of transitions $(0, a_j, s_{j+1}) \in (\Bf^{[j,j)})_{|\SSS_j \times \A_j \times \SSS_{j+1}}$.

\vspace{1ex}
\noindent
\textbf{Lemma}. For all $j$, $|(\CC_j)^c| = |(\SSS_j)^c| \cdot |(\Bf^{[j,j)})_{|\SSS_j \times \A_j \times \SSS_{j+1}}|.$

\vspace{1ex}
\noindent
\emph{Proof}.  The projection of $\Bf^c$ on $\SSS_j$ is $(\SSS_j)^c$, and its kernel is $\Bf^{[j,j)}$, so $(\SSS_j)^c \cong \Bf^c/\Bf^{[j,j)}$ by the FTH.
The projections of $\Bf^c$ and $\Bf^{[j,j)}$ on $\SSS_j \times \A_j \times \SSS_{j+1}$ are $(\CC_j)^c$ and $(\Bf^{[j,j)})_{|\SSS_j \times \A_j \times \SSS_{j+1}}$, respectively, and $\Bf^{[j,j+1)}$ is their common kernel, so $\Bf^c/\Bf^{[j,j)} \cong (\CC_j)^c/(\Bf^{[j,j)})_{|\SSS_j \times \A_j \times \SSS_{j+1}})$ by the CT.
  \qed \vspace{1ex}

Next, we define $P^c$  as the product of the sizes of all  controller granules up to level $n-1$, \ie
$P^c =  \prod_{\ell=0}^{n-1} \prod_{j\in\Z_n} |\Gamma^{[j,j+\ell]}|,$
 and $P = |\Gamma_\RR| P^c$ as the product of the sizes of all controller granules.  We observe that since $P$ is the number of possible sums of granule representatives, we have $|\Bf| \le P$, with equality if and only if  unique factorization holds for $\Bf$.  Similarly, we have $|\Bf^c| \le P^c$, with equality if and only if  unique factorization holds for $\Bf^c$.

\vspace{1ex}
\noindent
\textbf{Theorem} (\textbf{controllability and unique factorization}). Let $\Bf$ and $\Bf^c$ be the behaviors of a reduced linear or group tail-biting trellis realization $\RR$ and its controllable subrealization $\RR^c$, respectively.  Then: 

\vspace{1ex}
(a) $\RR^c$ is controllable. 

 (b) Unique factorization holds for $\Bf^c$;  \ie $|\Bf^c| = P^c$. 
 
  (c) $\RR$ is controllable if and only if  $\Bf = \Bf^c$;  \ie iff the top granule $\Gamma_\RR$ is trivial.
  
  (d) Unique factorization holds for $\Bf$;    \ie $|\Bf| = P$.

\vspace{1ex}
\noindent
\emph{Proof}. (a-b) From the previous lemma,  $\prod_j |(\CC_j)^c| = $ $ (\prod_j |(\SSS_j)^c|) (\prod_j |(\Bf^{[j,j)})_{|\SSS_j \times \A_j \times \SSS_{j+1}}|)$.  By  Section \ref{GI}, we have $|(\Bf^{[j,j)})_{|\SSS_j \times \A_j \times \SSS_{j+1}}| = \prod_{\ell=0}^{n-1} |\Gamma^{[j,j+\ell]}|$, so  
 $(\prod_j |(\CC_j)^c|)/(\prod_j |(\SSS_j)^c|)= \prod_j \prod_{\ell=0}^{n-1} |\Gamma^{[j,j+\ell]}|$ $ = P^c$, the product of the sizes of all proper controller granules $\Gamma^{[j,j+\ell]}$.  Therefore, by our controllability test, we  have $|\Bf^c| \ge  P^c$, with equality if and only if $\RR^c$ is controllable.
On the other hand,  in view of the controller granule decomposition of $\Bf^c$, we have $|\Bf^c| \le P^c$, with equality if and only if unique factorization holds for $\Bf^c$.  Thus $|\Bf^c| = P^c$,  $\RR^c$ is controllable, and unique factorization holds for $\Bf^c$. 

 (c)  By Section \ref{CTR},  $\Bf$ is the disjoint union of $|\Gamma_\RR|$ disconnected cosets of $\Bf^c$.  Thus we have $|\Bf| = |\Gamma_\RR| |\Bf^c|$, $|\CC_j| = |\Gamma_\RR| |(\CC_j)^c|$, and $|\SSS_j| = $ $ |\Gamma_\RR| |(\SSS_j)^c|$.  Therefore $(\prod_j |\CC_j|)/(\prod_j |\SSS_j|)$ $= P^c = |\Bf^c| = |\Bf|/|\Gamma_\RR|$. By our controllability test, $\RR$ is  controllable if and only if $|\Gamma_\RR| = 1$. 

 (d) By Section \ref{CTR}, every element of $\Bf$ is uniquely expressible as the sum of an element of $\Bf^c$ and a coset representative in $[\Gamma_\RR]$, so since unique factorization holds for $\Bf^c$, it  holds also for $\Bf$.
 \qed 
 
\subsection{State space and constraint code sizes}
 
 Unique factorization of $\Bf$ implies unique factorization of $\Bf_\FF$ for any fragment $\FF \le \RR$.  It follows that  the size of each state space $\SSS_j$ and each constraint code $\CC_j$ may be determined in terms of granule sizes as follows:
 
\vspace{1ex}
\noindent
\textbf{Corollary} (\textbf{state space and constraint code sizes}).  If $\RR$ is a reduced linear or group tail-biting trellis realization with state spaces $\SSS_j$ and constraint codes $\CC_j$, then: 

\vspace{1ex}
(a) $\SSS_j \cong \Bf/\Bf^{[j,j)}$, and 
$$|\SSS_j| =  \prod_{\FF \le \RR :~ \SSS_j \in E(\FF)} |\Gamma_\FF|;$$ 

(b)   $\CC_j \cong \Bf/\Bf^{[j+1,j)}$, and 
$$|\CC_j| = \prod_{\FF \le \RR :~ \CC_j \in V(\FF)} |\Gamma_\FF|.$$

\noindent
\emph{Proof}.  (a) If $\RR$ is state-trim at $\SSS_j$, then $\SSS_j = \Bf_{|\SSS_j}$.  Moreover,  the kernel of the projection of $\Bf$ onto $\SSS_j$ is  $\Bf^{[j,j)}$.   Thus $\SSS_j \cong \Bf/\Bf^{[j,j)}$ by the FTH, so $|\SSS_j| = |\Bf|/|\Bf^{[j,j)}| =  P/\prod_{\FF \le \FF^{[j,j)}} |\Gamma_\FF|=  \prod_{\FF \nleq \FF^{[j,j)}} |\Gamma_\FF| =  \prod_{\FF \le \RR \mid \SSS_j \in E(\FF)} |\Gamma_\FF|$, since  $\FF \le \FF^{[j,j)}$ iff $\SSS_j \notin E(\FF)$.  \vspace{1ex}

(b) If $\RR$ is branch-trim at $\CC_j$, then $\CC_j = \Bf_{|\SSS_j \times \A_j \times \SSS_{j+1}}$.  Moreover,  the kernel of the projection of $\Bf$ onto $\CC_j$ is  $\Bf^{[j+1,j)}$.   Thus $\CC_j \cong \Bf/\Bf^{[j+1,j)}$ by the FTH, so $|\CC_j| = |\Bf|/|\Bf^{[j+1,j)}| = P/\prod_{\FF \le \FF^{[j+1,j)}} |\Gamma_\FF| = \prod_{\FF \nleq \FF^{[j+1,j)}} |\Gamma_\FF| = \prod_{\FF \le \RR \mid \CC_j \in V(\FF)} |\Gamma_\FF|$, since  $\FF \le \FF^{[j+1,j)}$ iff $\CC_j \notin V(\FF)$.  
\qed \vspace{1ex}

In other words, assuming trimness, $\SSS_j$ factors into  components isomorphic to those granules $\Gamma_\FF$ such that $\SSS_j \in E(\FF)$ (\ie $\SSS_j$ is ``active" during $\FF$).  Also,   $\CC_j$ factors into  components isomorphic to those granules $\Gamma_\FF$ such that $\CC_j \in V(\FF)$ (\ie $\CC_j$ is ``active" during $\FF$).  

\subsection{Controller canonical  realization}\label{CCF}

The unique factorization result of Section \ref{CUF} implies that every reduced linear or group trellis realization is equivalent to a \emph{controller canonical  realization}, which we define as follows.  

For each $\FF \le \RR$, we have a one-to-one map $\Gamma_\FF \to  [\Gamma_\FF]$ from the  granule $\Gamma_\FF$ to the set of coset representatives $[\Gamma_\FF] = [\Bf_\FF/\Bf_{<\FF}]$.  We may thus map each element of the Cartesian product $\prod_{\FF \le \RR} \Gamma_\FF$ to the sum $(\ab, \sb) = \sum_{\FF \le \RR}   (\ab_\FF, \sb_\FF)$ of the corresponding coset representatives $(\ab_\FF, \sb_\FF) \in [\Gamma_\FF]$, which is an element of $\Bf$ since each coset representative is an element of $\Bf$.  By  unique factorization, the map so defined from $\prod_{\FF \le \RR} \Gamma_\FF$ to $\Bf$ is one-to-one.  

 \vspace{0.5ex}
More concretely,  the map $\prod_{\FF \le \RR} \Gamma_\FF \to \Bf$ may be implemented as follows. We generate the trajectories in $[\Gamma_\FF]$ by an \emph{atomic trellis realization} whose state spaces $\SSS_j$ are equal to $\Gamma_\FF$ when $\SSS_j \in E(\FF)$, and trivial otherwise.  An element of $\Gamma_\FF$ determines the state value $(\sb_\FF)_j$ when $\SSS_j \in E(\FF)$, and the symbol value $(\ab_\FF)_j$ when $\CC_j \in V(\FF)$.  The state value $\sb_j$ is thus the sum $\sum_{\FF \le \RR \mid \SSS_j \in E(\FF)} (\sb_\FF)_j$, and the symbol value $\ab_j$ is the sum $\sum_{\FF \le \RR \mid \CC_j \in V(\FF)} (\ab_\FF)_j$.
 The size of the aggregate state space $\SSS_j$  is thus $|\SSS_j| = \prod_{\FF \le \RR \mid \SSS_j \in E(\FF)} |\Gamma_\FF|$,  as in our state space size result.  Thus the controller canonical realization is a minimal  realization of  $\Bf$.  (We can also  show that the number of possible transitions $(s_j, a_j, s_{j+1})$ is $\prod_{\FF \le \RR \mid \CC_j \in V(\FF)} |\Gamma_\FF|$,  as in our constraint code  size result.) 
 
 \vspace{0.5ex}

If $\Bf$ is linear, then the controller canonical realization of $\Bf$ is easily seen to be linear.  However, for a group realization $\RR$, although the map $\prod_{\FF \le \RR} \Gamma_\FF \to \Bf$ yields a one-to-one, group-theoretic, and minimal realization of  $\Bf$, it may well not be isomorphic, even when $\RR$ is conventional \cite{FT93}.
This  issue was raised in \cite[Remark III.3]{C14} via the following  example, in which the   controller canonical realization is nonhomomorphic.  

\vspace{1ex}
\noindent
\textbf{Example} (Conventional group trellis realization over $\Z_4$).  Let $\RR$ be a conventional group trellis realization of length 3 with behavior $\Bf = \langle(112,0120)\rangle \subseteq (\Z_4)^3 \times (\Z_4)^4$;  \ie $\Bf = \{(000,0000), (112,0120),$ $(220, 0200), (332, 0320)\} \cong \Z_4$.  Its $\ell$-controllable subbehaviors are 
$\Bf_0 = \{(000,0000)\}$;
 $\Bf_1 = \Bf^{[0,1]} = \{(000,0000), (220, 0200)\} \cong 2\Z_4 \cong \Z_2$; and 
$\Bf_2 = \Bf \cong \Z_4$.
 Its nontrivial controller granules are $\Gamma^{[0,1]} = \Bf^{[0,1]} \cong \Z_2$, which is realized by a 2-state atomic trellis realization that is active during $[0,1]$, and $\Gamma^{[0,2]} = \Bf/\Bf^{[0,1]} \cong \Z_4/2\Z_4 \cong \Z_2$, which is realized by a 2-state atomic trellis realization that is active during $[0,2]$.  
 
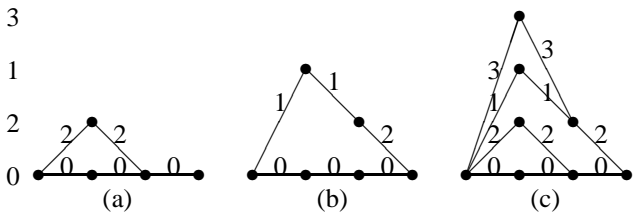
\begin{figure}[h]
\setlength{\unitlength}{4pt}
\centering
\begin{picture}(55,17)(0, -2)
\put(-3,-1){0}
\put(-3,9){1}
\put(-3,4){2}
\put(-3,14){3}
\put(0,0){\circle*{1}}
\put(0,0){\line(1,0){5}}
\put(0,0){\line(1,1){5}}
\put(2,0){0}
\put(2,3){2}
\put(5,0){\circle*{1}}
\put(5,5){\circle*{1}}
\put(5,5){\line(1,-1){5}}
\put(5,0){\line(1,0){5}}
\put(7,0){0}
\put(7,3){2}
\put(10,0){\circle*{1}}
\put(10,0){\line(1,0){5}}
\put(12,0){0}
\put(15,0){\circle*{1}}
\put(6,-3){(a)}

\put(20,0){\circle*{1}}
\put(20,0){\line(1,0){5}}
\put(20,0){\line(1,2){5}}
\put(22,0){0}
\put(22,6){1}
\put(25,0){\circle*{1}}
\put(25,10){\circle*{1}}
\put(25,10){\line(1,-1){5}}
\put(25,0){\line(1,0){5}}
\put(27,0){0}
\put(27,8){1}
\put(30,0){\circle*{1}}
\put(30,0){\line(1,0){5}}
\put(32,0){0}
\put(30,5){\circle*{1}}
\put(30,5){\line(1,-1){5}}
\put(32,3){2}
\put(35,0){\circle*{1}}
\put(26,-3){(b)}

\put(40,0){\circle*{1}}
\put(40,0){\line(1,0){5}}
\put(40,0){\line(1,1){5}}
\put(42,0){0}
\put(42,3){2}
\put(40,0){\line(1,2){5}}
\put(40,0){\line(1,3){5}}
\put(42,6){1}
\put(42,9){3}
\put(45,0){\circle*{1}}
\put(45,5){\circle*{1}}
\put(45,10){\circle*{1}}
\put(45,15){\circle*{1}}
\put(45,5){\line(1,-1){5}}
\put(45,0){\line(1,0){5}}
\put(45,15){\line(1,-2){5}}
\put(45,10){\line(1,-1){5}}
\put(47,0){0}
\put(47,7){1}
\put(47,11){3}
\put(47,3){2}
\put(50,0){\circle*{1}}
\put(50,0){\line(1,0){5}}
\put(52,0){0}
\put(50,5){\circle*{1}}
\put(50,5){\line(1,-1){5}}
\put(52,3){2}
\put(55,0){\circle*{1}}
\put(46,-3){(c)}
\end{picture}
\caption{Trellis diagrams for (a) $\Gamma^{[0,1]}$; (b) $[\Gamma^{[0,2]}]$;  (c) $\Bf$. \quad \qed}
\label{Ex3}
\end{figure}

 Figure \ref{Ex3} depicts the controller canonical realization of $\B$ via trellis diagrams for the atomic trellis realizations of $\Gamma^{[0,1]} = \Bf^{[0,1]}$ and $[\Gamma^{[0,2]}] = [\Bf/\Bf^{[0,1]}]$, plus a trellis diagram for $\Bf$.

\section{Conclusion}

We have generalized the CBFT to group trellis realizations, with a proof based on a controller granule decomposition of $\Bf$ and our controllability test for general group realizations.

It would be natural to dualize these results, using a dual observer granule decomposition.  However, as discussed in \cite{FT04}, such a dualization is not straightforward, even for minimal conventional trellis realizations.  Developing a nice dual observer granule decomposition for linear and group tail-biting trellis realizations is a good goal for future research.

It would  be nice also to extend these results to non-trellis realizations.  However, it is known (see \cite[Appendix A]{FGL12}) that unique factorization generally does not hold for non-trellis linear or group realizations, even  simple cycle-free realizations.  New ideas will therefore be needed.

Finally, we would like ultimately to redevelop all of the principal results of classical discrete-time linear systems theory using a purely group-theoretic approach.  However, the classical theory generally assumes an infinite time axis.  One possible approach would be to regard a  time-invariant or periodically time-varying linear or group system on an infinite time axis as the ``limit'' of a sequence of covers of a  linear or group tail-biting trellis realization on a sequence of finite time axes of increasing length.  Such an approach would hopefully be purely algebraic, and thus might avoid the subtle topological issues discussed in \cite{FT04}.  

\section*{Acknowledgments}

I am grateful to Mitchell Trott and Heide Gluesing-Luerssen for our earlier collaborations, in which many of these ideas first arose,  and to David Conti, for early access to and discussion of the results of \cite{C14}.

\end{document}